\documentclass[10pt,onecolumn,journal]{IEEEtran}

\usepackage{color}
\usepackage{graphicx}
\usepackage{graphics}
\usepackage{amscd}
\usepackage{amsmath}
\usepackage{amsfonts}
\usepackage{amssymb}
\usepackage{cite}
\usepackage{caption}

\makeatletter
\newcommand*\bigcdot{\mathpalette\bigcdot@{.5}}
\newcommand*\bigcdot@[2]{\mathbin{\vcenter{\hbox{\scalebox{#2}{$\m@th#1\bullet$}}}}}
\makeatother

\newtheorem{theorem}{Theorem}
\newtheorem{lemma}{Lemma}
\newtheorem{prop}{Proposition} 
\newtheorem{corollary}{Corollary}
\newtheorem{remark}{Remark} %number itself

\newtheorem{definition}{Definition}
\newtheorem{exam}{Example}

\newtheorem{con}{Construction}

\newenvironment{pf}{{\it Proof}.}{{\hfill $\square$%
        \hskip - \parfillskip}}

\begin{document}

\title{A New Centralized  Multi-Node Repair Scheme of MSR codes with Error-Correcting Capability}

\author{Shenghua Li,\, Maximilien Gadouleau,\,  Jiaojiao Wang,\, Dabin Zheng
     {\thanks{
   The work of Shenghua Li is in part funded by China Scholarship Council (No. 202108420166) . 
   \newline \indent Shenghua Li is with the Key Laboratory of Intelligent Sensing System and Secutity (Hubei Unibersity) , Ministry of Education, School of Cyber Science and Technology, Hubei University, Wuhan 430062, China (E-mail:  lish@hubu.edu.cn).
\newline \indent Maximilien Gadouleau (the corresponding author) is with Department of Computer Science, Durham University, Stockton Road, Durham DH1 3LE, UK (Email: m.r.gadouleau@durham.ac.uk). 
\newline \indent Jiaojiao Wang is  with the Data Science and Information Technology Research Center, Tsinghua-Berkeley Shenzhen Institute, Tsinghua
Shenzhen International Graduate School, Shenzhen 518052, China. (Email: wjj22@mails.tsinghua.edu.cn).
\newline \indent Dabin Zheng is with the Hubei Key Laboratory of Applied Mathematics, Faculty of Mathematics and Statistics, Hubei University, Wuhan 430062, China (Email: dzheng@hubu.edu.cn).
}}}

\date{ }

\maketitle

\begin{abstract}
    
Minimum storage regenerating (MSR) codes, with the MDS property and the optimal repair bandwidth, are widely used in distributed storage systems (DSS) for data recovery. In this paper, we consider the construction of $(n,k,l)$ MSR codes in the centralized model that  can repair 
$h$  failed nodes simultaneously with $e$ out $d$ helper nodes providing erroneous information.
We first propose the new repair scheme, and  give  a complete proof of the lower bound on the amount of symbols downloaded from the helped nodes, provided that some of helper nodes provide erroneous information.  Then we focus on two explicit constructions with the  repair scheme proposed. For $2\leq h\leq n-k$, $k+2e\leq d \leq n-h$ and $d\equiv k+2e \;(\mod{h})$, the first one has  the UER  $(h, d)$-optimal repair property, and the second one has the UER $(h, d)$-optimal access property. Compared with the original constructions  (Ye and Barg, \emph{IEEE Tran. Inf. Theory}, Vol. 63, April 2017), our constructions have improvements in three aspects: 1) The proposed repair scheme  is more feasible than  the one-by-one scheme presented by Ye and Barg  in a parallel data system;
2) The sub-packetization is reduced from $\left(\operatorname{lcm}(d-k+1,  d-k+2,\cdots, d-k+h)\right)^n$ to $\left((d-2e-k+h)/h\right)^n$, which reduces  at least by a factor of $(h(d-k+h))^n$;
3) The field size of the first construction is reduced to $|\mathbb{F}| \geq n(d-2e-k+h)/h$, which reduces  at least by a factor of $h(d-k+h)$. Small sub-packetization and small field size are preferred in practice due to the limited storage capacity and low computation complexity in the process of encoding, decoding and repairing.

\end{abstract}


% \linenumbers

\section{Introduction}

Owing to the optimal trade-off between the failure tolerance and storage overhead, maximum distance separable (MDS) codes are widely used in large-scale distribution storage systems (DSS). In such a system, a data file is stored across $n$ nodes, and the information of any $k$ nodes can reconstruct the original file (MDS property), i.e., the system can tolerate $r=n-k$ erasures. 
Though  having high storage efficiency, traditional MDS codes were illustrated to have low repair efficiency  by Dimakis et al. in \cite{Dimakis 2010}. Two important metrics for the repair efficiency are  the  amount of data downloaded and  the amount data accessed during the repair process, respectively. The former is called  {\it repair bandwidth}, indicating the network usage,  and the latter measures the disk input-output cost.  In \cite{Dimakis 2010}, the cut-set bound on repair bandwidth for a single failed node was derived, and regenerating codes were  defined as those achieving  the best trade-off between the repair bandwidth and storage overhead.  An important subclass of  regenerating codes is {\it minimum storage  regenerating} (MSR) code, which has the MDS property and the  optimal repair bandwidth.  Constructions of MSR codes were proposed in \cite{Dimakis 2010} -\cite{YeB 2017}, \cite{TamoWB 2013} -\cite{ZhouZ 2022}, and the references therein.

Multiple failed nodes are more frequent than  a single failed node  in practical DSS. Specially, in some systems,  the repair of erased nodes is only triggered once the number of failed nodes exceeds a determined threshold. Thus, it is usually desirable to repair multiple erasures efficiently  in DSS. There are two main models for repairing multi-node failures. One is the centralized model (CEM) where a repair center is assumed to reconstruct  all the  failed nodes (\cite{CadambeJMRS 2013} - \cite{MardiaBW 2019} ), and the repair bandwidth is the amount of data downloaded from the helped nodes for the repair. The other is the cooperative model (COM) where each  failed node downloads data from the helper nodes firstly, then they exchange information among themselves to finish the repair (\cite{ShumH 2013} - \cite{LiuCT 2023} ). Thus, the amount of data communicated among the failed nodes is also included in the repair bandwidth in the COM. The cut-set bounds  under these two models are obtained in \cite{CadambeJMRS 2013} and  \cite{ShumH 2013}, respectively. In this paper we only consider the centralized model, and will introduce the corresponding bounds later. In the digital network era, the nodes in DDS  are vulnerable to attacks from intruders. Since the intruders may be  the helper nodes, another basic repair issue of MDS codes is the case where information from some  helper
nodes is erroneous (\cite{YeB 2017}, \cite{PawarER 2011} -\cite{WangZLT 2023} ). Therefore,  it is of vital importance to study the repair for multi-node failures with error-correcting capability.

\vskip 0.3cm
{\noindent\it\large A. Cut-set Bounds}
\vskip 0.3cm

%2. The cut-set bound
MDS array codes \cite{Array Code}  are a special subclass of MDS codes that have  been extensively studied. An $(n, k, l)$ MDS array code over a field $\mathbb{F}$ can be viewed as a vector code with the MDS property, where  each coordinate of the codeword  is an $l$-dimension vector over $\mathbb{F}$. The parameter $l $ is called  the {\it sub-packetization }of the code, which is desired to be smaller in practice due to low complexity in the process of encoding and repairing.   Though being scalar MDS codes, Reed-Solomon (RS) codes can be viewed as vector codes over some subfield of $\mathbb{F}$, and the sub-packetization $l$ is defined as the degree of $\mathbb{F}$ over the subfield \cite{ShanPDC 2014}. The lower bound of sub-packetization for MSR codes is proved to be exponential (\cite{TamoYB 2019}, \cite{BalajiK 2018},\cite{AlrabiahG 2021} ).

%(h,d)-optimal repair property
Let  $\mathcal{C}=(C_1, C_2, \cdots, C_n)$ be an $(n, k, l)$ MDS array code over a finite field $\mathbb{F}$. Let $\mathcal{E}\subset [n]\; (=\{1,\cdots,n\}), |\mathcal{E}|=h$ and $\mathcal{R} = [n]\setminus {\mathcal{E}}, |\mathcal{R}| =d $ be the set  of indices of the failed nodes and the helper nodes, respectively.  Under the centralized model,  the repair center recover the values of the failed nodes by downloading $\beta_j, j\in \mathcal{R}$ symbols of $\mathbb{F}$ from each helper node $C_j, j\in \mathcal{R}$. Thus, the repair bandwidth is defined by
\begin{equation}\label{repair}
    \beta(\mathcal{E}, \mathcal{R})= \sum_{j\in \mathcal{R}}\beta_j.
\end{equation} 
The lower bound on the repair bandwidth is called the {\it cut-set} bound since it is obtained from the cut-set bound in network information theory. In \cite{Dimakis 2010}, \cite{CadambeJMRS 2013} and \cite{ RawatKV 2018}, the following inequality  for a single node and multiple nodes were derived respectively.
\begin{equation}\label{cut-bound1}
    \beta(\mathcal{E}, \mathcal{R}) \geq \frac{dhl} {h+d-k}.
\end{equation}
The $(h, d)$-repair bandwidth  of $\mathcal{C}$  (\cite {YeB 2017} ) is defined by 
$$\beta(h,d)= \max_{\mathcal{E}\cap\mathcal{R}=\emptyset, \mathcal{|E|}=h, \mathcal{|R|} =d}\beta(\mathcal{E}, \mathcal{R}).$$ 
If $\beta(h,d)$ meets the  bound  (\ref{cut-bound1}) with equality,  we say that $\mathcal{C}$  has the $(h, d)$-optimal repair property, and is called  an $(h, d)$-MSR code. 

%. UER

Suppose that  there exists erroneous information occurring in  at most $e$ out of  $d$ helper nodes. Let  $\beta(\mathcal{E, R},e)$   be the smallest amount of symbols of $\mathbb{F}$ downloaded from the  helper nodes $\{{C}_i, i\in \mathcal{R}\}$ to repair the failed nodes $\{{C}_i, i\in \mathcal{E}\}$  as long as the number of erroneous nodes in helper nodes is no more than $e$.  Define the universally error-resilient (UER) $(h, d)$-repair bandwidth of $\mathcal{C}$ as 
\begin{equation}\label{defhde}
\beta(h,d,e)= \max_{\mathcal{E}\cap\mathcal{R}=\emptyset, \mathcal{|E|}=h, \mathcal{|R|} =d }\beta(\mathcal{E}, \mathcal{R}, e).\end{equation} 
It was shown  in \cite{PawarER 2011}  and  \cite{RashmiSRK 2012} that $ \beta(1,d,e) \geq (dl)/(d-2e-k+1)$  for $d\geq k+2e$,  which  can be generalized for multiple failed nodes by a similar way. That is,
\begin{equation}\label{cut-bound2}
    \beta(h,d,e) \geq \frac{dhl} {d-2e-k+h},
\end{equation}
for any nonnegative integer $e$,   $h\geq 1$ 
 and $d\geq k+2e$.  If $\beta(h,d, e)$ meets the  bound  (\ref{cut-bound2}) with equality,  we say that $\mathcal{C}$  has the UER $(h, d)$-optimal repair property, and is called a UER $(h, d)$-MSR code.  To our knowledge, a complete proof of (\ref{cut-bound2}) was not given. Then, we will  restate it as  a theorem in next section,  and the  sufficient and necessary condition for   equality  will also be derived. 

In general, the amount of data accessed is larger than that of data downloaded in the repair process since the data downloaded may be a function of the data accessed. If the two are equal for an aforementioned MSR code, it is called more accurately $(h, d)$-optimal access code and UER $(h, d)$-optimal access code, respectively.

\vskip 0.3cm
{\noindent\it\large B. Related Work and  Our Contributions}
\vskip 0.3cm

%3. 
%known work   under the centralized model 

In many applications, it is desired to have a data centre responsible for the repair. For instance, in a rack-based system, there  exists a data  centre  in each rack which undertakes repairing the multiple-node failures in one rack (\cite{WangZLT 2023} - \cite{ChenB 2020} ). Furthermore,  The centralized repair model has applications in other communications, such as efficient secret sharing and broadcast (\cite{RawatKV 2018}, \cite{MitalKLG 2019} ).  \cite{CadambeJMRS 2013} first proposed the issue of repairing multi-node failures,  derived the lower bound of repair bandwidth and constructed the optimal codes by using
the asymptotic interference alignment (IA) technique. 
\cite{ZorguiW 2019} and \cite{RawatKV 2018} proposed the approaches to  constructions of MSR codes with multi-node failures in the CEM from known MSR codes with  a single node failure and cooperative MSR codes, respectively.  \cite{ZorguiW 2019} constructed  MSR codes with multi-node failures by using the  product-matrix (PM) technique. For  ZigZag (ZZ) codes (\cite{TamoWB 2013} ), \cite{WangTB 2017} studied the optimal repair for multi-node failures, and \cite{RawatKV 2018} discussed the special cases for two and three  node failures. Due to being used widely in practice, Reed-Solomon (RS) codes for multi-node failures have been extensively studied to seek a trade-off between the sub-packetization and the repair bandwidth.  \cite{TamoYB 2019} presented RS codes with optimal repair bandwidth for repairing multi-node failures in the CEM, and the sub-packetization $l=r!\prod_{i=1}^{n}p_i \approx n^n$, where $p_i$ is  the $i$-th smallest  prime  satisfying some properties. While taking the sub-packetization to be  on the order of $\log(n)$, \cite{DauDKM 2018} and \cite{MardiaBW 2019} proposed the repair schemes of RS codes with multi-node failures. 
In \cite{YeB 2017}, the authors considered  two constructions of UER $(h,d)$-MSR codes for all $h\leq r$ and $k\leq d\leq n-h$ simultaneously. Although not giving the explicit constructions, they proved that the amount of  data downloaded met the bound (\ref{cut-bound2}) under the one-by-one repair scheme. That is to say,   a failed node is first repaired by the helped nodes, then a second failed node is repaired by the helped nodes and the  first repaired node, then a third failed node is repaired  by the helped nodes and the first two repaired nodes , $\cdots$, finally he last remaining one is repaired  by the helper nodes and all the  previous repaired nodes. The sub-packetization of the codes is $\left(\operatorname{lcm}(d-k+1, d-k+2,\cdots, d-k+h)\right)^n$, which is at least $(d-k+h)^n(d-k+h-1)^n$. \cite{LiLL 2015} proposed a complete system supporting both single and concurrent failure
recovery,  and  showed that
the system achieves the minimum bandwidth for most concurrent failure patterns.

%5. 
In this paper, we study the simultaneous repair for multi-node failures, not the one-by-one repair. Repairing simultaneously means fairness for failed nodes. Moreover,  without dependencies, it is feasible for parallel processing, which will greatly improve the efficiency. Although there exist simultaneous repairs in the CEM, the sub-packetizations of MSR codes are very large (\cite {CadambeJMRS 2013}, \cite{TamoYB 2019} ),  huge finite field size is required for some of them ( \cite {CadambeJMRS 2013}, \cite{WangTB 2017}  due to the Schwartz–Zippel Lemma ), or just for some special $d$ (\cite{WangTB 2017} ), and they all have no consideration for the intruders. Large sub-packetization and huge finite field size can significantly increase the computational complexity of encoding and decoding. As far as we know, this paper is the first to focus on the simultaneous repair for multi-node failures with error-correcting capability. Comparisons of our MSR array codes  and some known MSR constructions with multi-node failures in the CEM are shown in Table \ref{tb1}, where $s=(d-2e-k+h)/h$ and  $s'={\rm lcm}(d-k+1,\cdots,d-k+h)$.  The main contributions of this paper are the proposed repair scheme and two constructions of MSR codes which have small sub-packetization and small field size. They are summarized as follows. 

1) We propose a repair scheme with error-correcting capability of MDS codes  for multi-node failures tolerance (see Definition \ref{def1}) and give a complete proof of the lower bound (\ref{cut-bound2}) on the repair bandwidth (see Theorem \ref{bound2}).  In the repair scheme, the process of repair  is  divided into $a$ groups. In each group, $l' = l/a$  coordinates in each failed node  are recovered,  and the total amount of  symbols downloaded from each helper node in all groups is  exactly $l/s$, which can ensure all coordinates in all failed nodes are obtained and the repair bandwidth meets the lower bound.  

2) Construction \ref{c1} gives a class of $(n,k,l=s^n)$ MDS array codes with  the UER $(h,d)$-optimal repair property (see Theorem \ref{orp2}), where $s=(d-2e-k+h)/h$ is an integer. The sub-packetization  and field size of codes are respectively set to be $s^n$, $sn$.  Compared with the construction in \cite {YeB 2017}, they reduce at least by factors of $(h(d-k+h))^n$ and $h(d-k+h)$, respectively.
The parity-check matrices of the codes are associated with diagonal  matrices, and another $(d-2e-k+n, d-2e, s^{n-h})$ MDS code is obtained for correcting $e$ errors. 

3) Construction \ref{c2} gives a class of $(n,k,l=s^n)$ MDS array codes with  the UER $(h,d)$-optimal access property (see Theorem \ref{UER2}), where $s=(d-2e-k+h)/h$ is an integer. The sub-packetization is set to be $s^n$, which  reduces at least by  a factor of $(h(d-k+h))^n$ compared with the construction in \cite {YeB 2017}. The parity-check matrices of the codes are associated with permutation  matrices, and another $(n-h, d-2e, s^{n-h})$ MDS code is obtained for correcting $e$ errors.

\begin{table}[!htb] 
    \centering
    \caption{\small Comparison of MSR codes for multi-node failures tolerance in the CEM}
    \renewcommand\arraystretch{1.3} 
    \footnotesize
    \begin{tabular}{|c|c|c|c|c|c|}
        \hline
        Construction & $l$  & Field size & $d$ &  UER &Scheme\\
        \hline
        IA $(n,k )$ Codes\cite{CadambeJMRS 2013} & $\begin {array} {c}
        (d-k+h)m^N, {\rm\; where}\\
        N = (d-k+h)(k-h),\\
         m \rightarrow \infty.\\
    \end{array}$ & large enough &  $k\leq d\leq n-h$  & NO& Simul.\\
    \hline
    
ZZ $(n,m+1 )$ Codes \cite{WangTB 2017} &  $r^m
$ & large enough  & $=n-h$ & NO&Simul.\\
\hline 
RS $(n, k )$ Codes \cite{TamoYB 2019} &  $\approx n^n $& very large: $p^l$ &  $k\leq d\leq n-h$  & NO&One.\\
\hline 
   $(n, k)$ Codes 1 \cite{YeB 2017}
     &  $(s')^n$ & $\geq ns'$ & $
    k+2e\leq d\leq n-h$  & YES &One.\\
    \hline 
   $(n, k)$ Codes 2 \cite{YeB 2017}
     &  $(s')^n$ & $\geq(n+1)$  & $
    k+2e\leq d\leq n-h$  &YES&One.\\
    \hline 
   $(n, k)$ Codes 1 our
     &  $s^n$ & $\geq ns$  & $\begin{array}{c}
    k+2e\leq d\leq n-h\\
    d\equiv k+2e \;(\mod{h})
\end{array} $  &YES&Simul.\\
    \hline 
   $(n, k)$ Codes 2 our
     &  $s^n$ & $\geq (n+1)$  & $\begin{array}{c}
    k+2e\leq d\leq n-h\\
    d\equiv k+2e \;(\mod{h})
\end{array} $  &YES&Simul.\\
\hline 
\end{tabular}\label{tb1}
\end{table}

\vskip 0.3cm
%4.
The rest of this paper is organized as follows. 
Section \ref{sec_pre} proposes the repair scheme firstly. Then two constructions of UER $(h,d)$-MSR codes for $2 \leq h\leq r$ and any corresponding reasonable $d$  are presented in Sections \ref{sec_c1} and \ref{sec_c2} respectively. Finally, Section \ref{sec_con} concludes this paper.

\section{Proposed Repair Framework  }\label{sec_pre}

In this section, we will illustrate and define the  repair scheme with error-correcting capability for multiple failures tolerance in the CEM, then give a complete proof to the lower bound  (\ref{cut-bound2}). For  ease of reading, we first introduce some notations used throughout in this paper.

\begin{itemize}
    \item  $\mathbb{F}$:  a finite field. \item $\mathbb{F}_{p^m}$: the finite field of   $p^m$ elements  for prime $p$ and integer $m\geq 1$. 
    \item $\mathbb{Z}_m$: the integer ring with $m$ elements.
    
     \item $[a, b] $:   the set  $\{a, a+1, \cdots, b\} $ with two integers $a$ and $b$, $a\le b$. If $a=1$,  $[b]$ is used in short. 
    
    \item   $n, k, r=n-k, l$: the number of total nodes, systematic nodes, parity nodes, and symbols in each node, respectively. 
    \item $h, d, e$: the number of failed nodes, total helper nodes, helper nodes with erroneous information  respectively.  Assume that $h\leq r$ and
    $k+2e\leq d\leq n-h$.
    \item  $s=(d-2e-k+h)/h$. For any $a\in [0, s^n-1]$, let $(a_{n}, a_{n-1}, \cdots, a_1)$ be its $s$-ary expansion form of length $n$. For integers $u_j\in [0, s-1]$, $j\in[n]$, $a(i_1, i_2, \cdots,i_j; u_1, u_2, \cdots,  u_j)$ denotes the $i_1$-th, $\cdots$,  $i_j$-th digits of $a$ are replaced with $u_1, u_2, \cdots,  u_j$, respectively.
    
    \item $\{e_a:  a= 0, 1, \cdots, l-1\}$:  the standard basis of $\mathbb{F}^l$ over $\mathbb{F}$.
    
    \item $|\Phi|$:  the cardinality of a set $\Phi$.
\end{itemize}

\vskip 0.3cm

\begin{figure}[htbp]
    \centering
    \includegraphics[width = 0.6\textwidth]{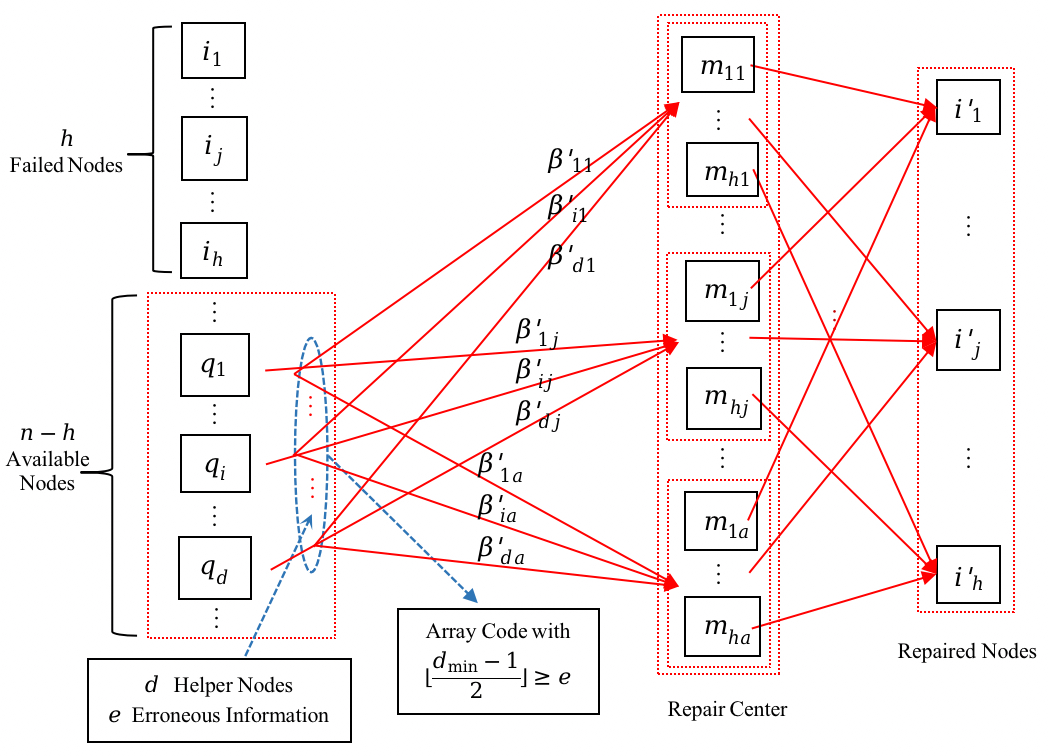}\\
    \caption{\small Repair for multiple-node failures simultaneously in the CEM:
        $h$ failed nodes, $e$ out of $d$ helper nodes with erroneous information}\label{Tcentralized}
\end{figure}

In the CEM, when $h$ nodes are failed, a repair
center is responsible for recovering the data stored in the failed nodes. If there exist  adversaries in the system, the errors in the helper nodes  must be corrected  firstly. Thus, we need to obtain another array code with the minimum distance $d_{\min}$ satisfying $\lfloor (d_{\min}-1)/2\rfloor \ge e$. Let $\mathcal{C}=(C_1, C_2, \cdots, C_n)$ be an $(n, k, l)$ MDS array code over  $\mathbb{F}$, where $C_i=(c_{i,0}, c_{i,1},\cdots, c_{i,l-1})$ is a vector of length $l$. Suppose the set of indices  of the failed nodes and helper nodes are $\mathcal{E}=\{i_1,\cdots, i_h\}\subset{[n]}$ and $\mathcal{R}=\{q_1,\cdots, q_d\}\subset{[n]}\setminus{\mathcal{E}}$, respectively.  For the sake of repairing all the failed nodes simultaneously, the coordinates to be recovered are divided into $a$ groups. In each group, $l' = l/a$ coordinates in each failed  node will be recovered by downloading the same amount of  symbols from each helper node. Provided  that  the information associated with  $d$ helper nodes will form a $(d, d-2e)$ array code, which can correct $e$ errors, the repair in each group are well done. After the repairs in all groups are finished,   $l$ coordinates in each node are obtained. 
Figure~\ref{Tcentralized} illustrates the process of the repair, where $\beta_{ij}' (i\in [d], j\in [a])$ represents the number of symbols downloaded from helper node $q_i$ in the $j$th group, and $m_{bj} (i_b\in [h], j\in [a])$ represents the set of symbols recovered of failed node $i_b$ in the $j$-th group. Thus, we have 
\begin{equation}\label{sum_m}\bigcup_{j=1}^{a}m_{bj} =  \{c_{i_b,0}, c_{i_b,1},\cdots, c_{i_b,l-1}\},  \end{equation} for all $b\in[h]$. We now mathematically formalize this repair scheme. 

\begin{definition}\label{def1}
    A  centralized repair scheme of $(n ,k,l)$ array code $\mathcal{C}$ over $\mathbb{F}$  with error-correcting capability for $h$-node  failures tolerance  is defined as follows. For each set $\mathcal{E}\subset{[n]}, |\mathcal{E}| = h$ and set $\mathcal{R}\subset{[n]}\setminus {\mathcal{E}}, |\mathcal{R}| = d$,
    \begin{itemize}
        \item [1)] The data associated with the  nodes $C_i, i\in \mathcal{R}$ forms a $(d, d-2e)$ array code. 
        \item [2)] There exist two positive integers $l', a$, s.t. $l=al'$, and functions 
        $$f_{i, j}:\; \mathbb{F}^{l}\rightarrow \mathbb{F}^{\beta'_{ij}}, \;\;i\in \mathcal{R},\; j\in [a]$$
        and $$g_j:\; \mathbb{F}^{\sum\limits_{i\in \mathcal{R}} \beta'_{ij}} \rightarrow \mathbb{F}^{hl'}, \; \;j\in [a]$$
        such that 
        $$\{(c_{i, 0}, c_{i, 1},\cdots, c_{i,l-1}): \; i\in \mathcal{E} \}= \bigcup_{j=1}^{a}g_j(  \{f_{i,j}((c_{i, 0}, c_{i, 1},\cdots, c_{i,l-1})), i\in \mathcal{R}\}).$$
    \end{itemize}
The repair bandwidth of this scheme is defined by
\begin{equation}\label{scheme}\sum_{j=1}^a \sum_{i\in \mathcal{R}}\beta'_{ij}.
\end{equation}
    \end{definition}

\vskip 0.3cm
Note that (\ref {scheme}) agrees with   (\ref{repair}),  here $\beta_i = \sum\limits_{j=1}^a \beta'_{ij}$, and the lower bound with error-correcting capability for multiple failures tolerance is given by  (\ref{cut-bound2}). For a single failure, the bound was discussed in  \cite{PawarER 2011} and \cite{RashmiSRK 2012}. For multiple failures,  (\ref{cut-bound2})  has been  stated (see \cite{YeB 2017}), but a complete proof is not given to our knowledge. Here, Theorem \ref{bound2} discuss the lower bound again and a complete proof is given based on the fact below. 
\begin{lemma} [\cite {WangZLT 2023} ] \label{WDLT}
    Let  $\mathcal{C}$ be an $(n, k, l)$ MDS array code over  $\mathbb{F}$. Suppose that $h$ failed nodes need to be repaired by $d$ helper nodes, $e$ of them providing erroneous information. Then, the total amount of  symbols over  $\mathbb{F}$ downloaded from any $d-2e-k+h$ helper nodes for the repair should be at least $hl$.
\end{lemma}
\begin{theorem}\label{bound2}
    Follow notations introduced  above. Then the minimum repair bandwidth for multi-node failures with error-correcting capability defined in (\ref{defhde}) satisfies (\ref{cut-bound2}). Moreover, the equality holds if and only if the number of symbols downloaded from each helper node is the  same, i.e.,  \begin{equation}
    \label{betai}
    \beta_i=(hl)/(d-2e-k+h), \end{equation}
    for all $i\in \mathcal{R}$.
\end{theorem}

\begin{pf}
    Let $\mathcal{I}\subset \mathcal{R}, |\mathcal{I}| = k+2e-h$ be a subset of helper nodes. Then $|\mathcal{R}\setminus \mathcal{I}|= d-2e-k+h$. By Lemma \ref{WDLT}, we have 
    \begin{equation}\label{Sum_betai}
        \sum_{i\in \mathcal{R} \setminus \mathcal{I}}\beta_i \geq hl,
    \end{equation}
    which $\beta_i$ is the number of  symbols downloaded from the node $i$. Summing the left-hand side over all $(k+2e-h)$-subsets of $\mathcal{R}$, we have 
    $$\sum_{\substack{\mathcal{I}\subset \mathcal{R} \\|\mathcal{I}| = k+2e-h}} \sum_{i\in \mathcal{R} \setminus \mathcal{I}}\beta_i 
    = \sum_{i\in \mathcal{R} }\sum_{\substack{\mathcal{I}\subset \mathcal{R} \\i\notin \mathcal{I}\\ |\mathcal{I}| = k+2e-h}}\beta_i 
    = \sum_{i\in \mathcal{R} }\binom{d-1}{k+2e-h}\beta_i
    = \binom{d-1}{k+2e-h}\sum_{i\in \mathcal{R} }\beta_i.$$
    Together with (\ref{Sum_betai}), we obtain
    $$\binom{d-1}{k+2e-h}\sum_{i\in \mathcal{R} }\beta_i \geq \binom{d}{k+2e-h}hl, $$
    which simplifies to (\ref{cut-bound2}).
    
    Next, we prove the second part. It is clear the equality in (\ref{cut-bound2}) holds if $\beta_i=(hl)/(d-2e-k+h)$  for each $i\in \mathcal{R} $. Note that the bound (\ref{cut-bound2}) holds with equality if and only if 
    (\ref{Sum_betai})  holds with equality for each $\mathcal{I}\subset \mathcal{R},\,  |\mathcal{I}| = k+2e-h$. Suppose that the equality in (\ref{cut-bound2})  holds with $\beta_{i^*} \neq (hl)/(d-2e-k+h)$ for some $i^*\in \mathcal{R}$, say $\beta_{i^*} < (hl)/(d-2e-k+h)$. Let $\mathcal{J}_1$ be a $(d-2e-k+h)$-subset of $\mathcal{R}$ containing $i^*$. Then, from $\sum_{i\in \mathcal{J}_1}\beta_i =  hl$, and there must $j_1 \in \mathcal{J}_1$, such that $\beta_{j_1} > (hl)/(d-2e-k+h)$.
    Let $\mathcal{J}_2$ be another $(d-2e-k+h)$-subset of $\mathcal{R}$ containing $j_1$ but not $i^*$. We also have  $\sum_{i\in \mathcal{J}_2}\beta_i  = hl$, and there must exists $j_2 \in \mathcal{J}_2, j_2\neq i^*$, such that $\beta_{j_2} < (hl)/(d-2e-k+h)$. Set $\mathcal{J}_3 = \{j_2\} \cup (\mathcal{J}_1 \setminus \{j_1\})$, then $\sum_{i\in \mathcal{J}_3}\beta_i  < hl$. With $|\mathcal{J}_3| = d-2e-k+h$, this contradicts (\ref{Sum_betai}), which completes the proof.
\end{pf}

\vskip 0.3cm
Since the data 
downloaded from the helper nodes may be  the same as the data accessed
on the helper nodes, the bound (\ref{cut-bound2}) is also a lower bound for the amount of  data accessed. Moreover,  the second construction in this paper shows that the lower bound is tight. 

\vskip 0.3cm
\begin{remark}
Note that  the repairs in  all groups can be carried out  at the same time if they are mutually independent. Since  the repair of each group in our scheme only relies on  helper nodes, our scheme can be applied in parallel processing. In \cite{YeB 2017}, the repair scheme is the one-by-one, and the failed nodes repaired  will be the helper nodes of the next failed node to be repaired. Thus,  the repair under  their scheme cannot be done in parallel.
\end{remark}

\begin{remark}
In this paper, $s=(d-2e-k+h)/h$ and $l=s^n$. From Theorem \ref{bound2}, the repair bandwidth bound is achieved if $\beta_i=l/s$ for each $i\in \mathcal{R}$, which coincide exactly with that in \cite{Dimakis 2010}. In the repair process of our array codes,  set $a = s^{h-1}$ and $\beta_{ij}'=s^{n-h}$ for all $i\in [d], j\in [a]$. Thus, $\beta_i=l/s$ for every $i\in \mathcal{R}$, and the codes are MSR codes.
\end{remark}
\begin{remark}
Note that (\ref{sum_m}) and (\ref{betai}) are the key to construct MSR codes with  error-correcting capability for multi-node failures tolerance. Now that any $d$ helper nodes will lead to a $(d,d-2e)$ array code, then $d$ helper nodes used in the repair of each group  can be different if (\ref{sum_m}) and (\ref{betai}) are also established. 
\end{remark}

\section {MSR Codes with the UER $(h,d)$-Optimal Repair Property }\label{sec_c1}

In this section, we first give the general array code construction  \cite{YeB 2017}  used in this paper. 
Let $\mathcal{C} \in \mathbb{F}^{ln}$ be an $(n, k, l)$ array code with nodes $C_i \in \mathbb{F}^{l}, i \in [n]$, where $C_i $ is a column vector $(c_{i, 0}, c_{i, 1},\cdots, c_{i, l-1})^T$. The code  $\mathcal{C}$ is defined by the following parity-check form:
\begin{equation}\label{array-code}
    \mathcal{C} = \{(C_1,C_2, \cdots, C_n): \;\sum_{i=1}^{n}A_{t,i} C_i=0,\;\; t\in [0, r-1]\},
\end{equation}
where $A_{t,i}$ is an $l\times l$ matrix  over $\mathbb{F}$ for  $t\in [0, r-1]$ and $i\in [n]$. The parity-check matrix in this paper is  a block Vandermonde matrix, i.e.,
\begin{equation}\label{parity-check}
    A_{t,i} = A_i^{t}, \;\; t \in [0, r-1], i \in [n], 
\end{equation}
where $A_i, i \in [n]$  are  $l\times l$ nonsingular matrices. By convention, $A^0 = I$ is used. 

We can characterize the  MDS property of $\mathcal{C}$ by its parity-check matrix. 
The code $\mathcal{C} $ defined by (\ref{array-code}) and (\ref{parity-check}) is MDS if and only if every $r\times r$ block submatrix of 
$$\left [
\begin{array}{cccc}
    I & I & \cdots &I\\
    A_1 & A_2 & \cdots &	A_n\\
    \vdots& \vdots & \vdots &\vdots \\
    A_1^{r-1} & A_2^{r-1} & \cdots &	A_n^{r-1}
\end{array}\right]$$
is invertible.  In \cite{YeB 2017}, Ye and Barg established the following criterion for invertibility of block matrices.
\begin{lemma} [\cite{YeB 2017} ] \label{Ven} Let $B_1, \cdots, B_r$ be $l\times l$ matrices such that $B_iB_j = B_jB_i$ for all $i, j\in [r]$. The matrix
    $$M_r = \left [
    \begin{array}{cccc}
        I & I & \cdots &I\\
        B_1 & B_2 & \cdots &	B_r\\
        \vdots& \vdots & \vdots &\vdots \\
        B_1^{r-1} & B_2^{r-1} & \cdots &	B_r^{r-1}
    \end{array}\right]$$
    is invertible if and only if $B_i-B_j$ is invertible for all $i \neq j$. 
\end{lemma}
\vskip 0.2cm

In this following, we discuss the simultaneous repair of multi-node failures for the MDS array codes associated with some diagonal matrices.
\begin{con}\label{c1}
    Let $\mathbb{F}$ be a finite field of size $|\mathbb{F}| \geq sn$. Let $\{\lambda_{i, j}\}_{i\in [n],\; j\in [0,s-1]}$ be $sn$ distinct elements in $\mathbb{F}$. Consider the  $(n, k , l = s^n)$ array code $\mathcal{C}$ defined by (\ref{array-code}) and (\ref{parity-check}), where  \begin{equation}\label{diag}
        A_i = \sum_{a=0}^{l-1}\lambda_{i,a_i}e_ae_a^T, \;\; i\in [n].\end{equation}
\end{con}
\vskip 0.3cm

From (\ref {diag}), we have $$A_i^t = \sum_{a=0}^{l-1}\lambda_{i,a_i}^te_ae_a^T, \; \; t\in [0,r-1],\;i\in [n].$$
Then, the parity-check equations (\ref {array-code}) coordinatewise can be rewritten as: 

\begin{equation}\label{pc1}
    \sum_{i=1}^{n}\lambda_{i,a_i}^tc_{i, a} = 0, 
\end{equation}
for all  $t \in [0, r-1], a\in [0, l-1]$. Namely,
$$\left [
\begin{array}{cccc}
    1 & 1 & \cdots &1\\
    \lambda_{1,a_1} & \lambda_{1,a_2} & \cdots &	\lambda_{1,a_n}\\
    \vdots& \vdots & \vdots &\vdots \\
    \lambda_{1,a_1}^{r-1} & \lambda_{1,a_2}^{r-1} & \cdots &	\lambda_{1,a_n}^{r-1}\\
\end{array}\right]
\left [
\begin{array}{c}
    c_{1,a}\\
    c_{2,a}\\
    \vdots \\
    c_{n,a}\\
\end{array}\right] = 0$$ 
for all $a\in [0, l-1]$.
To meet the MDS property,  it is required that every $r$ columns of the matrix above have rank $r$.  Since the submatrix composed of every $r$ columns is a Vandermonde matrix,  the code $\mathcal{C}$ in Construction \ref{c1} have the MDS property from $\lambda_{i, a_i}\neq \lambda_{j, a_j}$ for any $a \in [0, l-1]$ and $i\neq j$, $i,j\in [n]$. Then, we derive the optimal repair property of $\mathcal{C}$ as follows.

\vskip 0.2cm
\begin{theorem} \label{orp2}
    The code $\mathcal{C}$ given by Construction \ref{c1} has the UER $(h, d)$-optimal repair property for $k+2e\leq d\leq n-h$ and $d \equiv k+2e \;(\mod{h})$. 
\end{theorem}
\begin{pf}
    Let the set of indices of failed nodes be $\mathcal{E}=\{i_1, i_2,\cdots, i_h\}, 1\leq i_1<\cdots <i_h\leq n$.  For a fixed vector $(b_1, b_2, \cdots, b_{h-1})\in \mathbb{Z}_s^{h-1}$,   
    replacing $a$ with $a(i_1, \cdots, i_h; \;  u, u\oplus b_1, \cdots, u\oplus b_{h-1})$ in (\ref {pc1}),   we obtain
    
    \begin{equation} \label {bhu}
        \sum_{j=1}^h\lambda_{i_j, u\oplus b_{j-1}}^tc_{i_j, a(i_1,\cdots, i_h; \;  u, u\oplus b_1, \cdots, u\oplus b_{h-1})} = -\sum\limits_{i\in[n] \setminus \mathcal{E}}\lambda_{i, a_i}^t c_{i, a(i_1, \cdots, i_h;\;   u, u\oplus b_1, \cdots, u\oplus b_{h-1})}
        \end{equation}  
    for all  $t \in [0, r-1]$, where we set $b_0=0$.
    Summing the above over $u = 0, 1,\cdots , s - 1$,  we get 
    \begin{equation} \label{c1forh}
        \sum_{j=1}^h\sum\limits_{u=0}^{s-1}\lambda_{i_j, u\oplus b_{j-1}}^tc_{i_j, a(i_1, \cdots, i_h; \;  u, u\oplus b_1, \cdots, u\oplus b_{h-1})}   
        =-\sum\limits_{i\in[n] \setminus \mathcal{E}}\lambda_{i, a_i}^t\sum\limits_{u=0}^{s-1}c_{i, a(i_1, \cdots, i_h;\;   u, u\oplus b_1, \cdots, u\oplus b_{h-1})} , 
    \end{equation}
    for all $t\in [0, r-1]$.
    Note that $(hs+n-h)-r = d-2e$ and $ \{u\oplus b_i: u= 0, \cdots, s-1\}= [0, s-1]$ for any $b_i \in [0,s-1], i\in [h-1]$.  Then, (\ref{c1forh}) gives a $(d-2e-k+n, d-2e, s^{n-h})$ MDS array code $\mathcal{C'}$ from different $\lambda_{i,j}, i\in [n], j\in [0,s-1] $. Moreover, choosing any $d$ columns from  $\mathcal{C'}$ also constitutes a $(d, d-2e,s^{n-h})$ MDS array code, which can correct $e$ errors. Therefore, if we download any $d $  out of $n-h$ elements in the set
    $$\left \{\sum\limits_{u=0}^{s-1}c_{i, a(i_1, i_2,\cdots, i_h;\;   u, u\oplus b_1, \cdots, u\oplus b_{h-1})}:  \; i\in[n]\setminus \mathcal{E} \right\},$$ 
    we can recover the coordinates $\{a(i_1, i_2,\cdots, i_h; \;   u, u\oplus b_1, \cdots, u\oplus b_{h-1}):\; u=0,1\cdots, s-1\}$ of all the failed nodes as long as the number of erroneous nodes among the helper nodes is no more than $e$. When $(b_1, b_2, \cdots, b_{h-1})$ runs through all elements in $\mathbb{Z}_s^{h-1}$, all coordinates of the failed nodes are recovered. The amount of total downloaded data is $s^{h-1}\cdot d\cdot s^{n-h} = d\cdot s^{n-1} = dhl/(d-2e-k+h)$, which meets the bound (\ref{cut-bound2}). This completes the proof.
\end{pf}
\vskip 0.2cm
    From the right hand side of (\ref{c1forh}), the symbol downloaded from  each helper node for the repair is the sum of $s$ coordinates  of the node. Then, the amount accessed is $s$ times that of downloaded.
\begin{corollary}
For the code $\mathcal{C}$ given by Construction \ref{c1}, suppose that there exist at most $e$ of $d$ helper nodes providing erroneous information in the process of repairing $h$ failed nodes. Then, 
the total number of  symbols over  $\mathbb{F}$ accessed from helper nodes for the repair  is $dl$.
    
\end{corollary}

    \vskip 0.2cm
    \begin{exam} \label{Exam0}
        Let $(n, k, h, d, e) = (11, 3, 2, 7, 1)$,  then $r=8$ and $s=(d-2e-k+h)/h = 2 $. Let $\mathbb{F} =\mathbb{F}_{23}$,  and $\lambda_{1,0}, \lambda_{1,1}, \lambda_{2,0}, \lambda_{2,1}, \cdots, \lambda_{11,0}, \lambda_{11,1}$ be $22$ different elements of  $\mathbb{F}$. Consider the $(11, 3, 2^{11})$ array code  $\mathcal{C}$  over $\mathbb{F}$  defined by Construction \ref{c1}.   
        
        Assume that nodes 1, 2  are failed.  Let $b_1=0$ in (\ref{bhu}), we obtain 
        $$\lambda_{1, u}^tc_{1, a(1, 2; \;  u, u)}  + \lambda_{2, u}^tc_{2, a(1, 2; \;  u, u)}= -\sum\limits_{i=3}^{11}\lambda_{i, a_i}^t c_{i, a(1, 2;\;   u, u)}$$ for all $t\in [0, 7]$ and $u= 0, 1$. Then we get 
        $$(\lambda_{1, 0}^tc_{1, a(1, 2; \;  0, 0)} + \lambda_{1, 1}^tc_{1, a(1, 2; \;  1, 1)})+ (\lambda_{2, 0}^tc_{2, a(1, 2; \;  0, 0)}+\lambda_{2, 1}^tc_{2, a(1, 2; \;  1, 1)})= -\sum\limits_{i=3}^{11}\lambda_{i, a_i}^t (c_{i, a(1, 2;\; 0, 0)}+c_{i, a(1, 2;\; 1, 1)})$$ for all  $t\in [0, 7]$, which gives a $(13, 5, 2^9)$ MDS array code $\mathcal{C}_0$.  
         Then any $7$ columns of $\mathcal{C}_0$ also form a $(7, 5, 2^9)$ MDS array code, which can correct $e (=1)$ error. Thus, any $7$ columns in $\mathcal{C}_0$ can represent all columns in $\mathcal{C}_0$ as long as  the number of erroneous columns is no more than $e$.  So we obtain all  values in the set $$\{c_{1, (-,0,0)}, c_{1, (-,1,1)},\; c_{2, (-,0,0)}, c_{2, (-,1,1)} \}$$ 
        by downloading $2^{9}$ symbols from each helper node, where the symbol $'-'$ indicates that the upper 9 digits of the coordinate can take the value $0$ or $1$.
    
         Let $b_1 = 1$, another $(13, 5, 2^9)$ MDS array code $\mathcal{C}_1$ will be derived in  a similar way,  and all values in the set  $$\{c_{1, (-,0,1)}, c_{1, (-,1,0)},\; c_{2, (-,0,1)}, c_{2, (-,1,0)}\}$$ are also obtained by  downloading $2^{9}$ symbols  from each helper node.
         Thus, the repair bandwidth is $2\cdot 7\cdot 2^9 = 7\cdot 2^{10}$, which satisfies the lower bound (\ref{cut-bound2}). The total amount of symbols accessed is $ 7\cdot 2^{11}$.
   \end{exam}
\vskip 0.2cm
\begin{remark} For the same parameters $n, k, h,d$ in Example \ref{Exam0}, 
    the sub-packetization of MSR array codes in \cite{YeB 2017} is $$\left(\operatorname{lcm}(d-k+1, d-k+2)\right)^n = \left(\operatorname{lcm}(5, 6)\right)^n =(2\cdot 3\cdot 5)^{11}.$$
    The sub-packetization level of our array code   is reduced by a factor of $(3\cdot 5)^{11}$.  For $h=2, e=1$, our code at least reduces the sub-packetization  by a factor of $(2(d-k+2))^n$. Note that this factor will increase as $h$ or $e$ increases. Since $\left(\operatorname{lcm}(d-k+1,  d-k+2,\cdots, d-k+h)\right)^n\geq \left((d-k+h-1)( d-k+h)\right)^n$ for $h\geq 2$, our code reduces the sub-packetization in \cite{YeB 2017} at least by a factor of $(h(d-k+h))^n$. Moreover the size of $\mathbb{F}$ in  Example \ref{Exam0} is 23,  but the corresponding size of the field in \cite{YeB 2017}  is required to be at least $11\cdot (\operatorname{lcm}(5, 6)) = 330$. In general, the size of $\mathbb{F}$  in Construction \ref{c1}  is at least $h(d-k+h)$ times  less than that in \cite{YeB 2017}.
    Less sub-packetization and smaller  finite field  are better for MSR code in practice.
\end{remark}

\section {MSR Codes with the UER $(h,d)$-Optimal Access Property } \label{sec_c2}

In this section, we discuss the simultaneous repair  of multi-node failures for the MDS array code  associated with some permutation matrices. 

\begin{con}\label{c2}
    Let $\mathbb{F}$ be a finite field of size $|\mathbb{F}| \geq n+1$ and  $\gamma$  be a primitive element of $\mathbb{F}$. Consider the $(n, k , l = s^n)$ array code $\mathcal{C}$ defined by $(\ref{array-code})$ and $(\ref{parity-check})$, where the matrices are given by 
    \begin{equation}\label{perm} A_i = \sum_{a=0}^{l-1}\lambda_{i,a_i}e_ae_{a(i; a_i\oplus 1)}^T, \;\; i\in [n],\end{equation}
    where $\oplus$ denotes addition modulo $s$. 
    Here, $\lambda_{i, 0}=\gamma^i$ for all $ i\in [n]$ and $\lambda_{i, u}=1$ for all $ i\in [n]$ and all $u\in [s-1]$. 
\end{con}
\vskip 0.2cm

\vskip 0.2cm
From (\ref{perm}), we have 
$$A_i^t = \sum_{a=0}^{l-1}\beta_{i, a_i, t}e_ae_{a(i; a_i\oplus t)}^T,\; \; t\in [0,r-1], i\in[n]$$
where $\beta_{i,u,0} = 1$ and $\beta_{i,u,t} = \prod\limits_{v=u}^{u\oplus (t-1)}\lambda_{i,v}$ for $t \in [r-1]$ and $u \in [0,  s-1]$. Thus, the parity-check equations (\ref {array-code}) coordinatewise can be rewritten as: 

\begin{equation}\label{pc2}
    \sum_{i=1}^{n}\beta_{i, a_i, t}c_{i, a(i; a_i\oplus t)} = 0, \; \; {\rm for\; all\; } t \in [0, r-1], a\in [0, l-1].  
\end{equation}

One can check that $A_iA_j = A_jA_i$ and $A_i - A_j$ are invertible for any $i, j \in [n], i\neq j$. Then,   the code $\mathcal{C}$ given by  Construction  \ref{c2} is an MDS array code from Lemma \ref{Ven}.
\vskip 0.2cm

By a little computation, the following properties are obtained, which will be used for later computation.
$$A_i^s = \gamma^i I,\quad \prod\limits_{u=0}^{s-1}\lambda_{i,u}=\gamma^i\neq 1, \;i \in [n], $$

\begin{equation}\label{beta}
    \beta_{i,u,t} = \left\{\begin{array}{ll}
        1 & t=0, \\[3mm]
        \prod\limits_{v=u}^{u\oplus (t-1)}\lambda_{i,v} & t\in [s-1], \\ [3mm]
        \gamma^{j\cdot i}\beta_{i,u,t'} & t \in [s, r-1],\; j=\lfloor \frac{t}{s}\rfloor,\;t'=t-j\cdot s.
    \end{array}
    \right.
\end{equation}

\vskip 0.5cm

For Construction \ref{c2},  we first give an special example with $h=s=2, e=0$ and $d=n-h$ to illustrate the idea of our repair process.

\begin{exam} Let $n=6, k=h=2,  d=4$, and $e=0$. Then $r=4$ and $s=(d-k+h)/h =2$. By Construction \ref{c2},  we can obtain a $(6, 2, 2^6)$ MSR code over a finite field $\mathbb{F}$ with $|\mathbb{F}|\geq 7$.  Set $\mathbb{F} =\mathbb{F}_7$ and $\gamma = 3$. 
    
    Assume that nodes 1 and 2 are failed. Next, we show how to recover them. For some fixed $a\in [0, 2^6-1]$ with $ a_1=a_2=0$,  we  obtain four equations on the coordinates of the failed nodes by (\ref{pc2}) for  all $t\in [0,3]$ as follows.

    \begin {equation*}
    \left\{ \begin{array}{lllll}%{ccccc}
         c_{1, (-,0,0)}& +& c_{2, (-,0,0)}&=&-\sum\limits_{j=3}^{6}c_{j,a}\\[3mm]
        \gamma c_{1, (-,0,1)} &+& \gamma^2 c_{2, (-,1,0)}&=&-\sum\limits_{j=3}^{6}\gamma^{j} c_{j,a(j; a_j\oplus 1)}\\[3mm]
        \gamma c_{1, (-,0,0)} &+ &\gamma^2  c_{2, (-,0,0)}&=&-\sum\limits_{j=3}^{6}\gamma^{j} c_{j,a}\\[3mm]
        \gamma^2 c_{1, (-,0,1)} &+& \gamma^4 c_{2, (-,1,0)}&=&-\sum\limits_{j=3}^{6}\gamma^{2j} c_{j,a(j; a_j\oplus 1)}\\[5mm]
    \end{array}
    \right.
    \end {equation*}
    where the symbol $'-'$ denotes all other upper digits of the coordinate, and the coefficients are derived  from   (\ref{beta}). Then, by accessing the values in the set 
    $  \{c_{j, a}: \;  a_1=a_2=0, j\in[3,6]\}$, we can determine $c_{1, (-,0,0)}$ and $c_{2, (-,0,0)}$ by the 1st and 3rd equations, and determine  $c_{1, (-,0,1)}$ and $c_{2, (-,1,0)}$ by the 2nd  and 4th equations, since their coefficient matrices are invertible from $\gamma \neq \gamma^2$.
    Similarly,  $c_{1, (-,1,1)}$, $c_{2, (-,1,1)}$, $c_{1, (-,1,0)}$ and $c_{2, (-,0,1)}$ can be determined by accessing the values in the set 
    $  \{c_{j, a}: \;  a_1=a_2=1, \;j\in[3,6]\}$.
    
    Note that the parity-check equations for coordinate-elements in the set 
    \begin{equation}\label{set2}
        \{a:\; a_1=a_2=b,\;\; b=0,1\}
    \end{equation}
    lead to the two digits in the coordinates of nodes 1 and  2 running over $\mathbb{Z}_2^2$ exactly once. Thus, all symbols of nodes 1 and 2 can  be  obtained, and the number of symbols downloaded from four helped nodes is $4\cdot 2 \cdot 2^4 =(dhl)/(d-k+h)$, which achieves the bound (\ref{cut-bound1}). 
\end{exam}

\vskip 0.2cm
In general, we first show that the code $\mathcal{C}$ can be repaired optimally by downloading the data from the $n-h$ available nodes without error. Here, a subset of $\mathbb{Z}_s^h$ similar to (\ref{set2}) is required to ensure that all coordinates of the failed nodes are recovered and the repair bandwidth achieves the bound (\ref{cut-bound1}). 

Let \begin{equation}\label{Ahs}
    \Gamma(h, s) = \left\{(a_h,\cdots,a_2,a_1):\; \sum\limits_{i=1}^{h}a_i\equiv 0  \;(\mod{s}), \;\; a_i\in \mathbb{Z}_s, i\in[h]  \right\}.
\end{equation}

An element $(a_h,\cdots,a_2,a_1)$ in $\Gamma(h, s) $ corresponds to  a group in our repair scheme, and will be the sub-coordinate $(a_{i_h}, \cdots, a_{i_2}, a_{i_1})$ of $a$ in (\ref{cna1}). An important property of $\Gamma(h, s)$ is given as follows.

\begin{lemma}\label{lmAs}
    Let $\Gamma(h, s)$ be defined in (\ref{Ahs}). Then
    \begin{equation}\label{Zhs}
        \bigcup_{e\in \Gamma(h, s)}\bigcup_{t=0}^{s-1}e(i; e_i\oplus t) = \mathbb{Z}_s^h,
    \end{equation}
    for all $i \in [h]$.
\end{lemma}
\begin{pf}
    Note that  any two elements in $\Gamma(h, s)$ are different at least  at  two digits. Thus,  
    $$\left(\bigcup_{t=0}^{s-1}e(i; e_i\oplus t) \right)\bigcap \left (\bigcup_{t=0}^{s-1}e'(i; e'_i\oplus t)\right)=\emptyset,\quad e,e'\in \Gamma(h, s), \;e\neq e'$$
    for all $i\in [h]$. So the lemma holds from  $|\Gamma(h, s)|=s^{h-1}$ and $| \bigcup\limits_{t=0}^{s-1}e(i; e_i\oplus t)| = s$ for any $e\in \Gamma(h, s)$ and $i\in [h]$.
\end{pf}
\vskip 0.2cm
In the following, let  $\mathcal{E}=\{i_1, i_2, \cdots, i_h\}, 1\leq i_1<\cdots <i_h\leq n$ denote the indices of the $h$ failed nodes, where $2\leq h \leq r$. We rewrite  the parity equations (\ref{pc2}) as

\begin{equation}\label{cna1}
    \beta_{i_1, a_{i_1}, t} c_{i_1, a(i_1; a_{i_1}\oplus t)} +\cdots+ \beta_{i_h, a_{i_h}, t} c_{i_h, a(i_h; a_{i_h}\oplus t)}= -	\sum_{j\in[n]\setminus \mathcal{E} }\beta_{j, a_j, t} c_{j, a(j; a_j\oplus t)},
\end{equation}
for all $t \in [0, r-1], a\in [0, l-1]$. 
Let 
\begin{equation}\label{Bhs}
        B_{j, h, s}=\{c_{j, a}: (a_{i_h},\cdots, a_{i_2}, a_{i_1})\in \Gamma(h, s)\}, \; j\in [n].
    \end{equation} 
    We will show that all  coordinates of the failed nodes can be obtained by (\ref{cna1}) if the values in the set $B_{j, h,s},  j\in [n]\setminus \mathcal{E}$ are known.  From the right hand side of (\ref{cna1}), the number of symbols accessed will be  the same as that of symbols downloaded. 

\begin{lemma} \label{acph}
    Follow notations introduced above. The data in all failed nodes of the code $\mathcal{C}$ given by Construction \ref{c2} can be obtained simultaneously with the values in the set $B_{j, h,s},  j\in [n]\setminus \mathcal{E}$.
    
\end{lemma}
\begin{pf}
    Note that  $hs=d-2e-k+h\leq (n-h)-k+h=r$. From (\ref{cna1}),  we will obtain $hs$ equations with $t=0, 1, \cdots, hs-1$ for a fixed $a$, where   $b = (a_{i_h},\cdots, a_{i_2},a_{i_1})\in \Gamma(h, s)$. Moreover,
    for a fixed $p\in [0, s-1]$, the left hand sides of the $(j\cdot s+p)$-th equations, $j\in [0, h-1]$,  have the same one coordinate for each node $i$, $i \in\mathcal{E}$.  Thus, we can combine these $h$ linear equations to determine the $h$ unknowns since the coefficient matrix 
    $$\left [
    \begin{array}{cccc}
        x_1& x_2&\cdots&x_h\\
        \gamma^{i_1}x_1 & \gamma^{i_2}x_2&\cdots& \gamma^{i_h}x_h\\
        \gamma^{2\cdot i_1}x_1 & \gamma^{2\cdot i_2}x_2&\cdots& \gamma^{2\cdot i_h}x_h\\
        \vdots& \vdots&\cdots&\vdots\\
        \gamma^{(h-1)\cdot i_1}x_1 & \gamma^{(h-1)\cdot i_2}x_2&\cdots& \gamma^{(h-1)\cdot i_h}x_h\\
    \end{array}\right]$$  is invertible for all $\gamma^{i}\neq \gamma^{j}$, $i\neq j$, where $x_i, i\in [n]$ are nonzero (see (\ref{beta})). Thus, with the values in the set $\{c_{j, a}: (a_{i_h}, \cdots, a_{i_2},a_{i_1})=b, j\in [n]\setminus \mathcal{E}\}$,  we can get $s$ coordinates  of each node $i, i\in \mathcal{E}$ after solving $s$ groups of linear equations.
    From Lemma \ref {lmAs}, all coordinates of the failed nodes are obtained as  $b$ runs through all elements in the set $\Gamma(h, s)$. This completes the proof. 
\end{pf}
\vskip 0.2cm
Next, it remains to show that the code $\mathcal{C}$ has the UER $(h, d)$-optimal access property. It  is required to show that an $(n-h, d-2e, s^{n-h})$ MDS code is obtained. To this end,  we need the next lemma.  

\begin{lemma}  \label {lnoh}
    Follow notations introduced above. Then
    \begin{equation}\label{noh}
        \sum_{j\in [n]\setminus \mathcal{E}} \left(\prod_{k=1}^{h}(\gamma^{j}-\gamma^{i_k})   \right)A_j^m C_j = 0, 
    \end{equation}
    for all $m\in [0, r-hs-1].$
\end{lemma}
\begin{pf}
    Note that  $hs\leq r$ and $A_i^s = \gamma^i I$ for all $i\in [n]$. Then for any $p \in [0,h]$,
    \begin{equation}\label{Ap}
        \sum_{j=1}^n \gamma^{p\cdot j} A_j^m C_j  = \sum_{j=1}^n A_j^{m+p\cdot s} C_j = 0,
    \end{equation}
    for all $m\in [0, r-hs-1]$. By linearity, for any polynomial $P( x )$ of degree  $\leq h$, we have 
    \[
    \sum_{j=1}^n P( \gamma^j ) A_j^m C_j = 0.
    \]
    With $P(x) = \prod\limits_{k=1}^h ( x - \gamma^{i_k} )$, we obtain
    \[
    \sum_{j=1}^n \prod_{k=1}^h ( \gamma^j - \gamma^{i_k} ) A_j^m C_j = 0.
    \]
    Since the set of roots of $P(x)$ is exactly $\{\gamma^j, j\in\mathcal{E}\}$, we obtain the desired result.
\end{pf}

\begin{theorem} \label{UER2}
    The code $\mathcal{C}$ given by Construction \ref{c2} has the UER $(h,d)$-optimal access property for $h\leq r$, $ k+2e\leq d\leq n-h$ and $d\equiv k+2e \;(\mod{h})$.
\end{theorem}
\begin{pf}
    Note that $s=(d-2e-k+h)/h$ being an integer implies $d\equiv k+2e \;(\mod{h})$.
    From Lemma \ref{acph}, 
    we can repair the failed nodes if all values in the set $B_{j, h,s},  j\in [n]\setminus \mathcal{E}$ defined by (\ref{Bhs}) are known.
    
    Let $l' = s^{n-h}$. For a fixed $b\in \Gamma(h, s)$, define a function $f_b: [0, l'-1] \to [0, l-1]$ as  %$f_b: \{0, 1, \cdots,  l'-1\} \mapsto \{0, 1, \cdots, l-1\}$ as 
    $$f_b(a) = (a_{n-h},  \cdots, a_{i_h-h+1}, b_h, a_{i_h-h}, \cdots, a_{i_2-1}, b_{2}, a_{i_2-2}, \cdots, a_{i_1}, b_{1}, a_{i_1-1}, \cdots, a_1), $$
    where $a$ is an element in $[0, l'-1]$ with the $s$-ary expansion $(a_{n-h}, a_{n-h-1}, \cdots, a_1)$. For a fixed $b\in \Gamma(h, s)$, define the column vector $C_j^{(b, \mathcal{E})}\in \mathbb{F}^{l'}$ as 
    $$C_j^{(b, \mathcal{E})} = (c_{j, f_b(0)}, c_{j, f_b(1)},\cdots, c_{j, f_b(l'-1)})^T, $$
    for  all  $j\in[n] \setminus \mathcal{E}$. To prove the theorem, we only need to show that the vectors
    \begin{equation}\label{n-hc}
        \left(C_1^{(b, \mathcal{E})}, \cdots, C_{i_1-1}^{(b, \mathcal{E})}, C_{i_1+1}^{(b, \mathcal{E})}, \cdots, C_{i_h-1}^{(b, \mathcal{E})}, C_{i_h+1}^{(b, \mathcal{E})}, \cdots, C_{n}^{(b, \mathcal{E})}\right)
    \end{equation}
    form an $(n-h, d-2e, l')$ MDS array code for all $b\in \Gamma(h, s)$.
    
    Let $\{e_a^{(l')}:  a= 0, 1, \cdots, l'-1\}$ be  the standard basis of $\mathbb{F}^{l'}$ over $\mathbb{F}$. Define $l'\; \times\; l'$ matrices: 
    
    \begin{equation}\label{bj}
        B_j = \left\{\begin{array}{ll}
            \sum\limits_{a=0}^{l'-1}\lambda_{j, a_j }e_a^{(l')}(e_{a(j;  a_j\oplus 1)}^{(l')})^T,  & j \in [i_1-1] \\[4mm]
            \sum\limits_{a=0}^{l'-1}\lambda_{j+1, a_j }e_a^{(l')}(e_{a(j; a_j\oplus 1)}^{(l')})^T,  & j \in [i_1, i_2-2] \\[6mm]
            \sum\limits_{a=0}^{l'-1}\lambda_{j+2, a_j }e_a^{(l')}(e_{a(j; a_j\oplus 1)}^{(l')})^T,  & j \in [i_2-1, i_3-2] 
            \\[4mm]
            \cdots\\[4mm]
            \sum\limits_{a=0}^{l'-1}\lambda_{j+h-1, a_j }e_a^{(l')}(e_{a(j; a_j\oplus 1)}^{(l')})^T,  & j \in [i_{h-1}-1, i_h-2]
            \\[4mm]
            \sum\limits_{a=0}^{l'-1}\lambda_{j+h, a_j }e_a^{(l')}(e_{a(j; a_j\oplus 1)}^{(l')})^T,  & j \in [i_h-1, n-h]
            
        \end{array}
        \right.\end{equation}
    Thus,  (\ref{noh}) implies the following equations
    \begin{equation}\label{bpch}
        \begin{array}{ll}
            & \sum\limits_{j=1}^{i_1-1}\left(\prod\limits_{k=1}^{h}(\gamma^{j}-\gamma^{i_k})   \right) B_j^m C_j^{(b, \mathcal{E})} 
            +\sum\limits_{j=i_1}^{i_2-2}\left(\prod\limits_{k=1}^{h}(\gamma^{j+1}-\gamma^{i_k}) \right)B_j^m C_{j+1}^{(b, \mathcal{E})} \\[4mm]
            +&\sum\limits_{p=2}^{h-1}\;\sum\limits_{j=i_p-1}^{i_{p+1}-2}\left(\prod\limits_{k=1}^{h}(\gamma^{j+p}-\gamma^{i_k})\right)B_j^m C_{j+p}^{(b,\mathcal{E})} \\[4mm]
            +&\sum\limits_{j=i_h-1}^{n-h}\left(\prod\limits_{k=1}^{h}(\gamma^{j+h}-\gamma^{i_k})\right)B_j^m C_{j+h}^{(b,\mathcal{E})}  = 0
    \end{array}\end{equation}
    for all  $m\in [0, r-hs-1]$ and all $b\in \Gamma(h, s)$. 
    Next, we observe that $B_j$ is invertible for any $j \in [n-h]$, $B_{j_1}B_{j_2} =B_{j_2}B_{j_1}$, and $B_{j_1}-B_{j_2}$ is invertible  for any $j_1, j_2\in [n-h]$, $j_1\neq j_2$ (see Proposition \ref{Pbj2}  in Appendix \ref{ap1}).  Furthermore,   $r-hs = r- h\cdot (d-2e-k+h)/h = (n-h)-(d-2e)$. Thus, 
    $$\left [
    \begin{array}{cccc}
        I & I & \cdots &I\\
        B_1 & B_2 & \cdots &	B_{n-h}\\
        \vdots& \vdots & \vdots &\vdots \\
        B_1^{r-h\cdot s-1} & B_2^{r-h\cdot s-1} & \cdots &	B_{n-h}^{r-h\cdot s-1}
    \end{array}\right]$$
    is a parity-check matrix of an $(n-h,d-2e, l')$ MDS array code. Moreover, the coefficients before $B_j$ in (\ref{bpch}) are  nonzero for $ \gamma^x \neq \gamma^y, x\neq y $. Since multiplying each block column with a nonzero constant does not change the MDS property,  the vectors in (\ref {n-hc})  form an $(n-h, d-2e, l')$ MDS array code  $\mathcal{C}_b$, $b\in \Gamma(h, s)$. Therefore, if we access any $d$ out of $n-h$ vectors in this code, we can reconstruct the $n-h$ vectors  and further recover the $h$  failed nodes as long as the number of erroneous nodes among the helper nodes is no more than $e$.  The amount of total  data accessed is $d\cdot |\Gamma(h, s)|\cdot l'=d\cdot s^{n-1}=(d h l)/(d-2e-k+h)$, which meets the bound (\ref{cut-bound2}). This completes the proof.
\end{pf}
   
\vskip 0.3cm
\begin{exam} \label{Exam2}
Let $(n, k, h, d, e) = (15, 4, 3, 9, 1)$,  then $r=11$ and $s=(d-2e-k+h)/h = 2 $. Let $\mathbb{F} =\mathbb{F}_{2^4}$ and $\gamma$ be a primitive element of  $\mathbb{F}$. Consider the $(15, 4, 2^{15})$ array code  $\mathcal{C}$  over $\mathbb{F}$  defined by Construction \ref{c2}.   
    
    Assume that nodes 1, 2 and 3 are failed.  For some fixed $a\in [0, 2^{15}-1]$ with $ (a_3, a_2, a_1) \in \Gamma(3,2)$  (see (\ref{Ahs})), we  obtain six equations on the coordinates of the failed nodes by (\ref{pc2}) for $t \in [0, hs-1=5]$ as follows. (Here, we set $a=(-, 000)$, and `$-$' stands for all other binary bits)

    \begin {equation} \label{Ex2}
    \left\{ \begin{array}{lllllll}%{ccccccc}
        c_{1, (-,0,0,0)}& +& c_{2, (-,0,0,0)}&+&c_{3, (-,0,0,0)}&=&-\sum\limits_{j=4}^{15}c_{j,a}\\[3mm]
        \gamma c_{1, (-,0,0,1)} &+& \gamma^2c_{2, (-,0,1,0)}&+&\gamma^3 c_{3, (-,1, 0,0)}&=&-\sum\limits_{j=4}^{15}\gamma^j c_{j,a(j; a_j\oplus 1)}\\[3mm]
        \gamma c_{1, (-,0,0,0)} &+& \gamma^2c_{2, (-,0,0,0)}&+&\gamma^3 c_{3, (-,0, 0,0)}&=&-\sum\limits_{j=4}^{15}\gamma^j c_{j,a}\\[3mm]
        \gamma^2 c_{1, (-,0,0,1)} &+& \gamma^4c_{2, (-,0,1,0)}&+&\gamma^6 c_{3, (-,1, 0,0)}&=&-\sum\limits_{j=4}^{15}\gamma^{2j} c_{j,a(j; a_j\oplus 1)}\\[3mm]
        \gamma^2 c_{1, (-,0,0,0)} &+& \gamma^4c_{2, (-,0,0,0)}&+&\gamma^6 c_{3, (-,0, 0,0)}&=&-\sum\limits_{j=4}^{15}\gamma^{2j} c_{j,a}\\[3mm]
        \gamma^3 c_{1, (-,0,0,1)} &+& \gamma^6c_{2, (-,0,1,0)}&+&\gamma^9 c_{3, (-,1, 0,0)}&=&-\sum\limits_{j=4}^{15}\gamma^{3j} c_{j,a(j; a_j\oplus 1)}\\[5mm]
    \end{array}
    \right.
    \end {equation}
    where the coefficients are derived  from   (\ref{beta}). Then with the values in the set $\{c_{j, a}: \; a_1=a_2=a_3 = 0, \;j\in[4, 15] \}$, we can determine $c_{1, (-,0,0,0)}$, $c_{2, (-,0,0,0)}$, $c_{3, (-,0,0,0)}$ by the 1st, 3rd and 5th equations, and  determine $c_{1, (-,0, 0,1)}$, $c_{2, (-,0,1,0)}$ and $c_{3, (-,1,0,0)}$  by the 2nd, 4th and 6th equations, since their coefficient matrices are invertible from $\gamma \neq \gamma^2\neq \gamma^3$. 
    
    However, only $d=9$ helper nodes are set and one of them may provide the  erroneous information. To this end, we will show there exists another $(12, 7, 2^{12})$ MDS array code associated with all available nodes. Let $b=(0,0,0)=(a_3,a_2,a_1),\; \mathcal{E}=\{1,2,3\}$. Let $C_j^{(b,\mathcal{E})}=(c_{j, 0},\; c_{j, 8},\; \cdots,\; c_{j,2^{15}-8} )^T\in \mathbb{F}^{2^{12}}, j\in [4,15]$. Define  $\mathcal{C}_b = (C_4^{\;(b,\mathcal{E})},C_5^{\;(b,\mathcal{E})},\cdots, C_{15}^{\;(b,\mathcal{E})} )$. By the method in the proof of Theorem \ref{UER2}, $\mathcal{C}_b$ is proved to be an $(n-h, d-2e, s^{n-h}) = (12, 7, 2^{12})$ MDS array code. Then any $9$ columns of $\mathcal{C}_b$ also form a $(9, 7, 2^{12})$ MDS array code, which can correct $e (=1)$ error. Thus, any $9$ columns in $\mathcal{C}_b$ can represent all columns in $\mathcal{C}_b$ as long as  the number of erroneous columns is no more than $e$.  So by (\ref{Ex2}), we obtain the values  $\{c_{1, (-,0,0,0)}, c_{1, (-,0,0,1)},\; c_{2, (-,0,0,0)}, c_{2, (-,0,1,0)}, \; c_{3, (-,0,0,0)}, c_{3, (-,1,0,0)}\}$  
by downloading $s^{n-h}$ symbols from each helper node. 

Lemma \ref{lmAs} implies that all symbols of nodes 1, 2, and 3 can be obtained by taking the parity-check equations for all coordinate-elements in the set  $\{a:\; (a_3,a_2, a_1)\in \Gamma(3,2)\}$. The total number of symbols downloaded from helped nodes is $d \cdot s^{h-1}  s^{n-h} = 9 \cdot 2^{14}$, which achieves the bound (\ref{cut-bound2}). 
\end{exam}
\vskip 0.3cm
\begin{remark} For the same parameters $n, k, h,d$ in Example \ref{Exam2}, 
 the sub-packetization of MSR array codes in \cite{YeB 2017} is $$\left(\operatorname{lcm}(d-k+1, d-k+2, d-k+3)\right)^n = \left(\operatorname{lcm}(6, 7, 8)\right)^n = (3\cdot 7\cdot 2^3)^{15}.$$
The sub-packetization level of our array code   is reduced by a factor of $(3\cdot7\cdot 2^2)^{15}$. Reducing the sub-packetization of MSR codes is of great significance in practice. For $h=3, e=1$, our code at least reduces the sub-packetization  by a factor of $(3(d-k+2))^n$,  and $(3(d-k+2)(d-k+3))^n$ in the best scenario. 
\end{remark}

\section{Conclusion}\label{sec_con}
In this paper, a new explicit centralized repair scheme is proposed, and applied in two constructions of MSR codes associated with diagonal matrices and permutation matrices respectively. For $h$ failed nodes, $d$ helper nodes with $e$ adversaries, the sub-packetization $l$ is assigned $\left((d-2e-k+h)/h\right)^n$. When $h=1$, the codes are same as those codes for a single failed node in \cite{YeB 2017}. When $h\geq 2$, the sub-packetization of our codes is smaller than that of codes in \cite{YeB 2017}, where 
$l$ is $\left(\operatorname{lcm}(d-k+1,  \cdots, d-k+h)\right)^n$, and at least $(d-k+h)^n(d-k+h-1)^n$ for any $h\geq 2$. 

Unfortunately, since $(d-2e-k+h)/h$ is required to be an integer, our constructions only work for $d\equiv k+2e \;(\mod{h})$ helper nodes to repair $h$ failed nodes, and cannot work for an arbitrary $k+2e\leq d\leq n-h$. Therefore, our next task is to find an explicit centralized repair scheme for more admissible parameters $h, d$ with small  sub-packetization.

\begin{appendices}
\section{The properties of $B_j$ in (\ref{bj}) \label{ap1}}
\begin{prop}\label{Pbj2}
    Let $B_j, j\in[n-h]$ be defined by (\ref{bj}), then 
    \begin{itemize}
        \item [\romannumeral1)] $B_j$ is invertible for every $j \in [n-h]$, 
        \item [\romannumeral2)]  $B_{j_1}B_{j_2} =B_{j_2}B_{j_1}$ for any $j_1, j_2\in [n-h]$ and $j_1\neq j_2$, 
        \item [\romannumeral3)]  $B_{j_1}-B_{j_2}$ is invertible for any $j_1, j_2\in [n-h]$ and $j_1\neq j_2$.
    \end{itemize}
\end{prop}
\begin{pf}
    For easy clarifying the key idea of the proof, we only prove the properties in the case $h=2$. For other cases,  it can be proved in a similar way. 
    
     Let $n'=n-2$ and $l' = s^{n'}$. We first verify the invertibility of $B_j$ for every $j\in [n']$. According to (\ref{bj}), $|B_j|$ is equal to $\pm\gamma^{j\cdot l'/s}$, $\pm\gamma^{(j+1)\cdot l'/s}$ or $\pm\gamma^{(j+2)\cdot l'/s}$ depending on the   range  of  $j$, and nonzero. Thus, $B_j$ is invertible for every $j\in [n']$. Next, we verify the properties \romannumeral2) and \romannumeral3). Without loss of generality,  assume that $j_1\in [i_1-1]$ and $j_2\in [i_1, i_2-2]$. 
    
    For the commutativity, we have
    $$B_{j_1}B_{j_2} = B_{j_2}B_{j_1} = \sum\limits_{a=0}^{l'-1}\lambda_{j_1, a_{j_1} }\lambda_{j_2+1, a_{j_2}}e_a^{(l')}(e_{a(j_1, j_2; a_{j_1}\oplus 1, a_{j_2}\oplus 1)}^{(l')})^T. $$
    
    We now prove the invertibility of $B_{j_1} - B_{j_2}$. Suppose that $B_{j_1}x = B_{j_2}x$ for some vector $x\in \mathbb{F}^{l'}$. Let $\{e_a^{(l')}:  a= 0, 1, \cdots, l'-1\}$ be  the standard basis of $\mathbb{F}^{l'}$ over $\mathbb{F}$, and $x = \sum_{a=0}^{l'-1}x_ae_a^{(l')}$, where $x_a \in \mathbb{F}$. Then 
    $$B_{j_1}x = \sum_{a=0}^{l'-1}\lambda_{j_1, a_{j_1}}x_{a(j_1, a_{j_1}\oplus 1)}e_a^{(l')},\;B_{j_2}x = \sum_{a=0}^{l'-1}\lambda_{j_2+1, a_{j_2} }x_{a(j_2, a_{j_2}\oplus 1)}e_a^{(l')}.$$
    Therefore, 
    \begin{equation}\label{diff}
        \lambda_{j_1, a_{j_1}}x_{a(j_1, a_{j_1}\oplus 1)} = \lambda_{j_2+1, a_{j_2} }x_{a(j_2, a_{j_2}\oplus 1)}\end{equation}for all $a\in [0, l'-1]$. since $\lambda_{i, u}\neq 0$ for all $i \in [n]$ and $u \in [0,s-1]$, we can rewrite (\ref{diff}) as
    $$x_a = \frac{\lambda_{j_2+1, a_{j_2} }}{\lambda_{j_1, a_{j_1} \ominus 1}}x_{a(j_1, j_2; a_{j_1} \ominus 1, a_{j_2}\oplus 1)}$$ 
    Repeating this operation, we obtain 
    $$\begin{array}{lll}
        x_a &=&\frac{\lambda_{j_2+1, a_{j_2} }}{\lambda_{j_1, a_{j_1} \ominus 1}}\cdot \frac{\lambda_{j_2+1, a_{j_2} \oplus 1 }}{\lambda_{j_1, a_{j_1} \ominus 2}} x_{a(j_1, j_2; a_{j_1} \ominus 2, a_{j_2}\oplus 2)}= \cdots\\[5mm]
        & = &  \frac{\lambda_{j_2+1, a_{j_2} }}{\lambda_{j_1, a_{j_1} \ominus 1}}\cdot \frac{\lambda_{j_2+1, a_{j_2} \oplus 1 }} {\lambda_{j_1, a_{j_1} \ominus 2}} \cdots \frac{\lambda_{j_2+1, a_{j_2} \oplus (s-2) }} {\lambda_{j_1, a_{j_1} \ominus (s-1)}}\cdot \frac{\lambda_{j_2+1, a_{j_2} \oplus (s-1) }} {\lambda_{j_1, a_{j_1} \ominus s}} x_{a(j_1, j_2; a_{j_1}   \ominus s, a_{j_2}\oplus s)}\\[5mm]
        & = &\frac{\prod_{u=0}^{s-1}\lambda_{j_2+1, a_{j_2} \oplus u }}{\prod_{u=0}^{s-1}\lambda_{j_1, a_{j_1} \oplus u }} x_a = \lambda^{j_2+1-j_1}x_a
    \end{array}$$
    for all $a\in [0,l'-1]$. Since $\lambda^{j_2+1-j_1} $ is not equal to $0$ or $1$, we have $x_a = 0$ for all $a\in [0,l'-1]$. Thus, $(B_{j_1}-B_{j_2})x = 0$ implies $x = 0$, and we can draw a conclusion that $B_{j_1} - B_{j_2}$ is invertible.  This completes the proof.
\end{pf}
\end{appendices}

\end{document}